\documentclass[twocolumn,english,prb,notitlepage]{revtex4-1}
\usepackage[T1]{fontenc}
\usepackage[latin9]{inputenc}
\usepackage{babel}
\usepackage{float}
\usepackage{amsmath}
\usepackage{graphicx}
\usepackage[unicode=true,pdfusetitle,
 bookmarks=true,bookmarksnumbered=false,bookmarksopen=false,
 breaklinks=false,pdfborder={0 0 1},backref=false,colorlinks=false]
 {hyperref}

\usepackage{dsfont}

\usepackage{csquotes}
\MakeOuterQuote{"}

\makeatletter

\providecommand{\tabularnewline}{\\}

\usepackage{xcolor}
\usepackage{caption}

\@ifundefined{showcaptionsetup}{}{%
 \PassOptionsToPackage{caption=false}{subfig}}
\usepackage{subfig}
\makeatother

\begin{document}

\newcommand{\br}{\mathbf{r}}
\newcommand{\bF}{\mathbf{F}}
\newcommand{\E}{\mathcal{E}}
\newcommand{\G}{\mathcal{G}}
\newcommand{\Lag}{\mathcal{L}}
\newcommand{\M}{\mathcal{M}}
\newcommand{\N}{\mathcal{N}}
\newcommand{\U}{\mathcal{U}}
\newcommand{\R}{\mathcal{R}}
\newcommand{\F}{\mathcal{F}}
\newcommand{\V}{\mathcal{V}}
\newcommand{\D}{\mathcal{D}}

\newcommand{\ca}{c^{\phantom{\dagger}}}
\newcommand{\cc}{c^\dagger}
\newcommand{\fa}{f^{\phantom{\dagger}}}
\newcommand{\fc}{f^\dagger}
\newcommand{\aaa}{a^{\phantom{\dagger}}}
\newcommand{\aac}{a^\dagger}
\newcommand{\bba}{b^{\phantom{\dagger}}}
\newcommand{\bbc}{b^\dagger}
\newcommand{\da}{{d}^{\phantom{\dagger}}}
\newcommand{\dc}{{d}^\dagger}
\newcommand{\ha}{h^{\phantom{\dagger}}}
\newcommand{\hc}{h^\dagger}
\newcommand{\be}{\begin{equation}}
\newcommand{\ee}{\end{equation}}
\newcommand{\bea}{\begin{eqnarray}}
\newcommand{\eea}{\end{eqnarray}}
\newcommand{\ba}{\begin{eqnarray*}}
\newcommand{\ea}{\end{eqnarray*}}
\newcommand{\dagga}{{\phantom{\dagger}}}
\newcommand{\bR}{\mathbf{R}}
\newcommand{\bQ}{\mathbf{Q}}
\newcommand{\bq}{\mathbf{q}}
\newcommand{\bqp}{\mathbf{q'}}
\newcommand{\bk}{\mathbf{k}}
\newcommand{\bh}{\mathbf{h}}
\newcommand{\bkp}{\mathbf{k'}}
\newcommand{\bp}{\mathbf{p}}
\newcommand{\bL}{\mathbf{L}}

\newcommand{\Pj}[2]{|#1\rangle\langle #2|}
\newcommand{\ket}[1]{\vert#1\rangle}
\newcommand{\bra}[1]{\langle#1\vert}
\newcommand{\setof}[1]{\left\{#1\right\}}
\newcommand{\fract}[2]{\frac{\displaystyle #1}{\displaystyle #2}}
\newcommand{\Av}[2]{\langle #1|\,#2\,|#1\rangle}
\newcommand{\Avs}[1]{\langle \,#1\,\rangle_0}
\newcommand{\eqn}[1]{(\ref{#1})}
\newcommand{\Tr}{\mathrm{Tr}}

\title{Charge self-consistent density functional theory plus ghost rotationally-invariant
slave-boson theory for correlated materials}
\author{Tsung-Han Lee$^{1,2}$, Corey Melnick$^{3}$, Ran Adler$^{1}$, Xue Sun$^{1}$, Yongxin
Yao$^{4}$, Nicola Lanat\`a$^{5,6}$, Gabriel Kotliar$^{1,3}$}

\affiliation{$^{1}$Physics and Astronomy Department, Rutgers University, Piscataway,
New Jersey 08854, USA}
\affiliation{$^{2}$Department of Physics, National Chung Cheng University, Chiayi 62102, Taiwan}
\affiliation{$^{3}$Condensed Matter Physics and Materials Science Department,
Brookhaven National Laboratory, Upton, New York 11973, USA}
\affiliation{$^{4}$Ames National Laboratory and Iowa State University, Ames, Iowa
50011, USA}
\affiliation{$^{5}$School of Physics and Astronomy, Rochester Institute of Technology,
84 Lomb Memorial Drive, Rochester, New York 14623, USA}
\affiliation{$^{6}$Center for Computational Quantum Physics, Flatiron Institute,
New York, New York 10010, USA}
\begin{abstract}
We present a charge self-consistent density functional theory combined with the ghost-rotationally-invariant slave-boson (DFT+gRISB) formalism for studying correlated materials. This method is applied to SrVO$_3$ and NiO, representing prototypical correlated metals and charge-transfer insulators. For SrVO$_3$, we demonstrate that DFT+gRISB yields an accurate equilibrium volume and effective mass close to experimentally observed values. Regarding NiO, DFT+gRISB enables the simultaneous description of charge transfer and Mott-Hubbard bands, significantly enhancing the accuracy of the original DFT+RISB approach. Furthermore, the calculated equilibrium volume and spectral function reasonably agree with experimental observations.
\end{abstract}
\maketitle

\section{Introduction}

Simulating strongly correlated materials from first
principles remains one of the most formidable challenges
in condensed matter physics. The complexities arise
from the intricate interplay among electronic charge, spin, and orbital degrees of freedom, as well as the electron's dual localization and itinerancy character in these materials,
driven by strong local Coulomb interactions. This necessitates the use of quantum many-body techniques that go beyond standard \textit{ab initio} density functional theory (DFT)  \cite{Hohenberg1964,Kohn1965} for their description.

The combination of DFT with dynamical mean field theory (DFT+DMFT) has been extraordinarily successful in addressing this challenge~\cite{Anisimov_Kotliar_1997,Lichtenstein_Kotliar_2001}. The DMFT, as the first example of a quantum embedding approach, maps the interacting lattice to an auxiliary quantum impurity model with self-consistently determined bath orbitals~\cite{DMFT_RMP_1996}, allowing an accurate description of the local correlation physics. Moreover, the DFT+DMFT has been extended to charge self-consistency deriving from a functional formulation~\cite{DMFT_RMP_2006,Held2007}.  The method requires a suitable selection of a correlated set of orbitals~\cite{Savrasov2004,Pavarini_2004,Pourovskii_2007,korotin_2008,Haule_DFT+DMFT_2010}, the value of the interaction parameters~\cite{Anisimov_1993_slater_condon,Anisimov_1997_slater_condon}, a suitable double-counting correction~\cite{FLL,Anisimov_1997_slater_condon,Haule_2015_exactDC}, and accurate impurity solvers~\cite{Gull_RMP_2011,Bulla2008}.
This framework is now well-developed, and comparisons between experiments and theory have revealed new physics in many correlated materials, shedding light on phenomena such as Mott localization~\cite{Koga_OSMT,Medici2005,Yu_2013,Deng2019}, Hund's physics~\cite{Georges_Hunds_review,Medici_PRB_2011,Medici_Janus_PRL,Medici_instability_2017,Werner_power_law,HauleHunds}, and the valence fluctuations in correlated systems~\cite{Savrasov2006,Shim2007,Marc2015}. 
Nevertheless, the approach is computationally demanding and sometimes suffers from the so-called sign problem in the Quantum Monte Carlo solver with sizable off-diagonal hybridizations~\cite{Gull_RMP_2011}.
 
Another approach starts from the Gutzwiller approximation (GA)~\cite{Gutzwiller1,Gutzwiller2,Metzner_Vollhardt_1989,Bunemann_mulorb_GA,Bunemann_GA_RISB,Fabrizio_SC,Lanata_2008,Lanata_2012} and equivalently rotationally-invariant slave-boson (RISB) method~\cite{Lecherman_2007,Lanata_2017_PRL}, and their combination with DFT~\cite{Ho_2008,Deng_2008,Deng_2009,Yongxin_2011_1,Yongxin_2011_2,Christoph_2011,Lanata_2013_Ce,Lanata_2015_PRX,Lanata_2015_PRX,Lanata_2017_PRL}.
These methods, realized as quantum embedding approaches~\cite{Lanata_2015_PRX,gDMET}, similar to DMFT, map the lattice problem onto an embedded impurity model and are connected to other quantum embedding concepts~\cite{DMET_2012,Lanata_2015_PRX,Ayral_DMET_RISB,RISB_DMET_Lee_2019,gDMET}.
The RISB framework can capture local Mott, Hund's, and valence fluctuation physics~\cite{Yongxin_2011_1,Lanata_2012,Lanata_2013_Ce,Lanata_2015_PRX,Lanata_2017_PRL}, at a lower computational cost compared to DMFT. However, it sometimes suffers from insufficient accuracy, particularly failing to capture the interplay
between the Mott physics and charge fluctuations~\cite{gRISB_2021}.

The ghost-rotationally-invariant-slave-boson (gRISB) method was recently introduced to overcome these limitations, expanding the RISB variational space by employing auxiliary ``ghost'' fermionic degrees of freedom~\cite{gRISB_2017,gRISB_2021,gRISB_2022}. 
Studies have shown that gRISB, even with a small number of ghost orbitals, consistently achieves ground-state and spectral properties that closely align with DMFT, across both single- and multi-orbital Hubbard models~\cite{gRISB_2017,gRISB_2021,Lee_one-orbital_2023,Guerci2019,Carlos2023,Lee_multiorb_2023}. Additionally, numerical evidence indicates that the accuracy of gRISB approaches that of DMFT solutions as the number of ghost bath orbitals is increased~\cite{Guerci_thesis,Lee_one-orbital_2023,Guerci2019,Lee_multiorb_2023}. This accuracy was confirmed through direct comparisons with DMFT, using exact diagonalization as an impurity solver and discretized hybridization functions~\cite{Caffarel_Krauth_1994}.
Moreover, the gRISB requires the calculation of only the ground-state single-particle density matrix of the embedding Hamiltonian, avoiding the need to compute dynamic quantities of the impurity model, making it computationally efficient. The gRISB also does not require a bath fitting procedure and can be seamlessly combined with the density matrix renormalization group (DMRG) solvers~\cite{White1992,White1993,Schollwock2005,Lee_multiorb_2023} and machine learning methods~\cite{Rogers_ML,Marius_ML_2024}.
These features position gRISB as a promising approach warranting further investigation, particularly in combination with DFT.

In this work, we present a charge self-consistent DFT plus gRISB (DFT+gRISB) formalism to simulate correlated materials. We apply DFT+gRISB to SrVO$_3$ and NiO, representing correlated metal and charge-transfer insulator systems, respectively, and compare the results with DFT+DMFT. For SrVO$_3$, we demonstrate that DFT+gRISB yields reliable total energy and mass renormalization, in good agreement with experiments and DFT+DMFT studies, significantly improving upon the original RISB approach. For NiO, we show that DFT+gRISB provides a consistent description of the charge-transfer insulator,
consistent with experimental and DFT+DMFT studies, while DFT+RISB falsely predicts a metallic solution for NiO. The DFT+gRISB total energy is also in good agreement with DFT+DMFT. 
Finally, we demonstrate the applicability of the DMRG solver within the gRISB framework, allowing for accurate results, including full five d-orbitals. 

\section{Method}

In this section, we discuss the formalism of the charge-self-consistent DFT+gRISB approach and the implementation of our DFT+gRISB framework.

\subsection{Formalism}

The DFT+gRISB functional is encoded in a Lagrange function~\cite{Lanata_2015_PRX,DMFT_RMP_2006}
represented as follows:
\begin{align}
&\mathcal{L}^\text{DFT+gRISB}_N\big[
\rho(\br),\mathcal{J}(\br),\mu,
V_i^0,N_i^0
\big]=\mathcal{L}_\text{gRISB}\big[\mathcal{J}(\br),\mu\big]
\nonumber
\\
&\quad 
-\int \! d\br \,\rho(\br) \mathcal{J}(\br)
+E_{\text{Hxc}}[\rho(\br)]
+E_{\text{ion-ion}}+E_{\text{ion}}[\rho(\br)]
\nonumber
\\
&\quad 
+ \sum_{i}E^i_{\text{dc}}\left[N^0_i\right] - \sum_i V_i^0 N^0_i
+\mu (N + \sum_i m_i)
\label{LN}
\,,
\end{align}
where $N$ is the total number of electrons in the system, determined by the charge-neutrality condition, $\mu$ is the chemical potential, $\rho(\br)$ is the electron density, $\mathcal{J}(\br)$ is the corresponding constraining field, $E_{\text{Hxc}}$ is the Hartree exchange-correlation functional, $E_{\text{ion-ion}}$ is the ion-ion energy, $E_\text{ion}$ is the ionic potential, 
$E_{dc}^i\left[N_i^0\right]=\frac{U_i}{2}N_i^0\left(N_i^0-1\right) - \frac{J_i}{2} N_i^0 \left(\frac{N_i^0}{2} - 1\right)$  
is the double-counting energy functional associated with the $i$-th impurity~\cite{FLL,Anisimov_1997_slater_condon}, $N^0_i$ is the corresponding occupancy, $V_i^0$ is the corresponding potential, and $U_i$ and $J_i$ is the corresponding Coulomb interaction and Hund's coupling interaction, respectively.

The term $\mathcal{L}_\text{gRISB}$ is the gRISB Lagrange function associated with following many-body Kohn-Sham-Hubbard "reference system," expressed in second quantization as follows:
\begin{align}
    &\hat{H}_{\text{KSH}}=
    \int\! dx\, \hat{\Psi}^{\dagger}\left(x\right)
    \hat{P}
    \left[-\hat{\nabla}^{2}+\mathcal{J}\left(\hat{x}\right)-\mu\right]
    \hat{P}\,
    \hat{\Psi}\left(x\right)
    \nonumber
    \\
    &\quad+
    \sum_{\bR, i}\left(
    \hat{H}_i^{\text{int}}[\cc_{\bR i\alpha},\ca_{\bR i\alpha}]
    +V_i^0\sum_\alpha \cc_{\bR i\alpha}\ca_{\bR i\alpha}
    \right)
    \label{KSH-LAPW}
    \,,
\end{align}
where $x=(\mathbf{r},\sigma)$, $\sigma$ is the spin variable, $\mathbf{r}$ is the position variable, $\int dx$
indicates both the sum over $\sigma$ and the integral over $\mathbf{r}$,
$\nabla^2$ is the Laplacian, $\hat{\Psi}(x)$ is the Fermionic field operator, $\hat{P}$ is the projector over a generic computational basis span and:
\begin{align}
\ca_{\bR i\alpha}&=\int \! dx \, 
\phi^*_{\bR i \alpha}(x)\hat{\Psi}(x)
\\
\phi_{\bR i \alpha}(\br)&=\mathcal{N}^{-1}
\sum_{\bk}e^{-i\bk\bR}\phi_{\bk i \alpha}(x)
\end{align}
are the annihilation operators associated with the corresponding correlated degrees of freedom, where 
$\bR$ is the unit cell label, $\bk$ is the momentum, 
and $\mathcal{N}$ is the total number of unit cells in the system.
The correlated orbital function is denoted by $\phi_{\bR i \alpha}(x)$, where $\alpha=1,..,\nu_i$ encodes both the orbital degrees of freedom and the spin.

The Lagrangian $\mathcal{L}_\text{gRISB}$ can be formally expressed as follows:
\begin{align}
&\mathcal{L}_\text{gRISB}\big[\mathcal{J}(\br),\mu\big]=
-\frac{T}{\mathcal{N}}\sum_\omega\text{Tr log}\left[i\omega - H^\text{qp}\right]
\nonumber\\
&+\mathcal{N}\sum_{i}\Big[\langle\Phi_{i}|\hat{H}^{\text{emb}}|\Phi_{i}\rangle+E_{i}^{c}\big(1-\langle\Phi_{i}|\Phi_{i}\rangle\big)\Big]\nonumber \\
 & -\mathcal{N}\sum_{i}\Big[\sum_{ab}\big(\big[\lambda_{i}\big]_{ab}+\big[\lambda_{i}^{c}\big]_{ab}\big)\big[\Delta_{i}\big]_{ab}\nonumber \\
 & +\mathcal{N}\sum_{ac\alpha}\big(\big[D_{i}\big]_{a\alpha}\big[R_{i}\big]_{c\alpha}\big[\Delta_{i}(1-\Delta_{i})\big]_{ca}^{\text{\ensuremath{\frac{1}{2}}}}+\text{c.c.}\big)\Big]
\,,
\end{align}
where $H^{\text{qp}}$ is the single-particle matrix representation of the so-called quasiparticle Hamiltonian:
\begin{align}
\hat{H}^{\text{qp}}&=
    \int\! dx\, \hat{\Psi}_{\text{u}}^{\dagger}\left(x\right)
    \hat{P}
    \left[-\hat{\nabla}^{2}+\mathcal{J}\left(\hat{x}\right)-\mu\right]
    \hat{P}
    \,
    \hat{\Psi}_{\text{u}}\left(x\right)
    \nonumber\\
    &+\int\! dx\, \hat{\Psi}_{\text{c}}^{\dagger}\left(x\right)
    \hat{P}
    \left[-\hat{\nabla}^{2}+\mathcal{J}\left(\hat{x}\right)-\mu\right]
    \hat{P}
    \,
    \hat{\Psi}_{\text{c}}\left(x\right)
    \nonumber\\
    &+\Big(
    \int\! dx\, 
    \hat{\Psi}_{\text{c}}^{\dagger}\left(x\right)
    \hat{P}
    \left[-\hat{\nabla}^{2}+\mathcal{J}(\hat{x})\right]
    \hat{P}
    \,
    {\Psi}_{\text{u}}\left(x\right)
    \nonumber\\
    &
    +\text{H.c.}
    \Big)+
    \sum_{\bk i}\sum_{ab}
     \left[\lambda_i-\mathcal{E}_i^{\text{qp}}\right]_{ab} f_{\bk i a}^\dagger f_{\bk i b}
    \label{hqp-2quant}
\end{align}
for a given computational basis projector $\hat{P}$ and correlated orbital wavefunction $\phi_{\bR i\alpha}(x)$, and the local correlated part of the $H^{\text{qp}}$ has the form:
\begin{equation}
[\mathcal{E}^{\text{qp}}_i]_{ab} = \sum_{\alpha\beta}
    [R_i]_{a\alpha} [\mathcal{E}^{\text{loc}}_i]_{\alpha\beta}  [R^\dagger_i]_{\beta b}
\end{equation}
with 
\begin{equation}
[\mathcal{E}^\text{loc}_i]_{\alpha\beta} =
    \frac{1}{\mathcal{N}}\sum_\bk
    \langle\phi_{\bk i\alpha}|-\hat{\nabla}^2+\mathcal{J}(\hat{x})|\phi_{\bk i\beta}\rangle.
    \label{Eloci}
\end{equation}
The matrix elements of $R_i$ and $\lambda_i$ are the so-called renormalization coefficients of the quasiparticle Hamiltonian, $\Delta_i$ is the quasiparticle single-particle density matrix.
The $\hat{{\Psi}}_{\text{u}}(x)$ and $\hat{\Psi}_{\text{c}}$ is the uncorrelated and correlated part of the field operator, respectively, defined as follows:
\begin{align}
\hat{\Psi}_{\text{u}}(x)&=
\left[
\hat{I}
    -\sum_{\bk i \alpha}
    \ket{\phi_{\bk i \alpha}}
    \bra{\phi_{\bk i \alpha}}
\right]
\hat{\Psi}(x)
\\
\hat{\Psi}_{\text{c}}(x)&=
\sum_{\bk i a} {f}_{\bk i a}
\sum_\alpha
[R^\dagger_i]_{\alpha a}
\phi_{\bk i \alpha}(x)
\,,
\end{align}
where $\hat{I}$ is the identity operator, $a=1,..,B\nu_{i}$, with $B$ controlling the accuracy of the gRISB method, 
and $f_{\bk ia}$ are the so-called quasi-particle annihilation operators. We have also introduced the so-called embedding Hamiltonian of the $i$-th impurity:
\begin{align}
    \hat{H}^{\text{emb}}_i &= \hat{H}_i^{\text{int}}[c^\dagger_{i\alpha},c^\dagga_{i\alpha}]
    +V_i^0\sum_\alpha c^\dagger_{i\alpha}c^\dagga_{i\alpha}
    \nonumber \\
    &+\sum_{\alpha\beta} 
    [\mathcal{E}^\text{loc}_i]_{\alpha\beta}
    c^\dagger_{i\alpha}c^\dagga_{i\beta}
    +\sum_{a\alpha}\left([D_i]_{a\alpha}\, 
    c^\dagger_{i\alpha}b^\dagga_{ia}
    +\text{c.c.}\right)
    \nonumber \\ 
    &+\sum_{ab} [\lambda^c_i]_{ab}\, 
    b^\dagga_{ib}b^\dagger_{ia}
    \,.
\end{align}
The matrix elements of
$D_i$ and $\lambda_i^c$ is the hybridization and bath coupling constants, respectively, $\ket{\Phi_i}$ is the ground state of $\hat{H}^{\text{emb}}_i$, $E^c_i$ is a Lagrange multiplier enforcing the normalization of $\ket{\Phi_i}$, and
$c_{i\alpha}$, $b_{ia}$ are the impurity and bath Fermionic annihilation operators, respectively. The number of spin orbitals in the bath is $N_{b,i}=B\nu_i$.

The charge neutrality is enforced by the chemical potential $\mu$ at quasiparticle occupancy $N+\sum_{i}m_{i}$, where $m_{i}=(B\nu_{i}-\nu_{i})/2$.
The reason for this additional term $m_{i}$ is to enforce the physical occupancy to be at the total physical valence number $N$, where the quasiparticle occupancy and the physical occupancy differs by a number $\sum_{i}m_{i}$~\cite{gRISB_2021,gRISB_2022}.
The $\lambda_i$ and $\lambda_i^c$ can also be viewed as the Lagrange multiplier enforcing the gRISB constraints, and $D_i$ is a Lagrange multiplier enforcing the structure of the $\left[R_{i}\right]_{,a\alpha}=\langle\Phi_i|c^\dagger_{i\alpha} b_{ib}|\Phi_i\rangle\big[\Delta_i(1-\Delta_i)\big]_{ba}^{-\frac{1}{2}}$ matrix~\cite{Lecherman_2007}.

The stationary condition of the DFT+gRISB functional leads to the following saddle-point equations: 
\setlength{\jot}{0.15cm}
\begin{widetext}
\begin{gather}
\mathcal{J}(\mathbf{r})=\frac{\delta H_{\text{Hxc}}^{\text{LDA}}\big[\rho(\mathbf{r})\big]}{\delta\rho(\mathbf{r})}+\frac{\delta E_{\text{ion}}\big[\rho(\mathbf{r})\big]}{\delta\rho(\mathbf{r})},\label{eq:sp1}
\\
\frac{1}{\mathcal{N}} \sum_{\bk}\langle f_{\bk i a}^\dagger f_{\bk i b} \rangle_{0}=\left[\Delta_{i}\right]_{ab},\label{eq:sp3}
\\
\rho(\br)  =\langle \hat{\Psi}_{\text{u}}^\dagger(\br) \hat{\Psi}_{\text{u}}(\br)\rangle_{0}
+ \langle \hat{\Psi}_{\text{c}}^\dagger(\br)\hat{\Psi}_{\text{c}}(\br) \rangle_{0} +\left(\langle \hat{\Psi}_{\text{c}}^\dagger(\br) \hat{\Psi}_{\text{u}}(\br)\rangle_{0} + \text{H.c.}\rangle\right)\qquad\qquad\nonumber \\
\qquad\qquad+\frac{1}{\mathcal{N}}\sum_i\sum_{\mathbf{k}}\sum_{\alpha\beta}\phi^*_{\bk i \alpha}(\br)\left(\langle \Phi_{i}|c^\dagger_{i\alpha} c_{i\beta}|\Phi_i\rangle - \sum_{ab} [R_i^\dagger]_{\alpha a} \left[\Delta_{i}\right]_{ab} [R_i]_{b\beta} \right)\phi_{\bk i \beta}(\br),\label{eq:sp2}
\\
\int dx \frac{1}{\mathcal{N}}\sum_{\mathbf{k}} \sum_{b}\sum_{i'} \phi^*_{\bk i \alpha}(x)\hat{P}\left[-\nabla^2+J(\hat{x})-\mu\right]\hat{P}\phi_{\bk i'\beta}(x) [R_i^\dagger]_{\beta b} \langle f_{\bk ia}^\dagger f_{\bk i' b}\rangle_{0} \qquad \quad \nonumber \\
\qquad\quad+\int dx \frac{1}{\mathcal{N}}\sum_{\mathbf{k}} \phi^*_{\bk i \alpha}(x)\hat{P}\left[-\nabla^2+J(\hat{x})\right]\hat{P}\langle f_{\bk ia}^\dagger \hat{\Psi}_{\text{u}}(x)\rangle_T=\sum_{c}\left[D_{i}\right]_{c\alpha}\big[\Delta_{i}(1-\Delta_{i})\big]_{ac}^{\frac{1}{2}},\label{eq:sp4}
\\
\int dx \left[\langle\hat{\Psi}_{\text{u}}^\dagger(x) \hat{\Psi}_{\text{u}}(x)\rangle_{0}
+ \langle \hat{\Psi}_{\text{c}}^\dagger(x)\hat{\Psi}_{\text{c}}(x) \rangle_{0}\right] =N+\sum_{i}m_{i},\label{eq:sp5}
\\
\sum_{cd\alpha} \frac{\partial}{\partial d_{i,s}}\Big(\big[\Delta_{i}(1-\Delta_{i})\big]_{cd}^{\frac{1}{2}}\left[D_{i}\right]_{d\alpha}\left[R_{i}\right]_{c\alpha}+\text{c.c.}\Big)+l_{i,s}+l^c_{i,s}=0,\label{eq:sp6}
\\
\hat{H}_{i}^{\text{emb}}|\Phi_{i}\rangle  =E_{i}^{c}|\Phi_{i}\rangle,\label{eq:sp7}
\\
\langle\Phi_{i}|c_{i\alpha}^{\dagger}b_{ia}|\Phi_{i}\rangle-\sum_{c}\big[\Delta_{i}(1-\Delta_{i})\big]_{ac}^{\frac{1}{2}}\left[R_{i}\right]_{c\alpha}=0,\label{eq:sp8}
\\
\langle\Phi_{i}|b_{ib}b_{ia}^{\dagger}|\Phi_{i}\rangle-\left[\Delta_{i}\right]_{ab}=0,\label{eq:sp9}
\end{gather}
\end{widetext}
where $\langle...\rangle_0$ denotes the average over the ground state of $H^{\text{qp}}$, and we introduced the following parameterization of the matrices:
\vspace{1cm}
\newpage
\begin{align}
\left[\Delta_i\right]_{ab} &= \sum_{s} d_{i,s} \left[h_{i,s}\right]_{ab} \\
\left[\lambda_i\right]_{ab} &= \sum_{s} l_{i,s} \left[h_{i,s}\right]_{ab} \\
\left[\lambda^c_i\right]_{ab} &= \sum_{s} l^c_{i,s} \left[h_{i,s}\right]_{ab},
\end{align}
where $h_{i,s}$ is an orthonormal basis of the Hermitian matrices~\cite{Lanata_2017_PRL}. Equation \ref{eq:sp1} gives rise to the Kohn-Sham potential, and Eq. \ref{eq:sp2} is the DFT+gRISB charge density, where the local quasiparticle contribution to the density is subtracted and replaced with the contribution from the local physical density matrix~\cite{Deng_2009}. The chemical potential is determined from Eq. \ref{eq:sp5}.
The other equations are the standard gRISB equations for model Hamiltonian~\cite{gRISB_2017,gRISB_2021,gRISB_2022,Carlos2023,Lee_multiorb_2023}.
The detailed algorithm for solving these equations will be discussed in the next subsection.

With the converged $R_i$ and $\lambda_i$, one can compute the Green's function as follows: 
\begin{align}
\big[&G_i(\mathbf{k},\omega)\big]_{\alpha\beta} \nonumber\\ 
&=\sum_{ab}[R^\dagger_i]_{\alpha a}\langle0|  f_{\bk ia} \frac{1}{\omega+i\eta -\hat{H}^{\text{qp}}+\mu} f^{\dagger}_{\bk i b }  |0\rangle \left[R_i\right]_{b\beta}
\,,\label{eq:G}
\end{align}
where $|0\rangle$ is the vacuum. Equation~\ref{eq:G} holds because the $\hat{H}^{\text{qp}}$ is a single-particle Hamiltonian. The spectral function is calculated from 
$A_i(\bk,\omega)=-\text{Im}G_i(\bk,\omega)/\pi$,
and we use a broadening factor of $\eta=0.05$ eV. 

The self-energy can be determined from the Dyson equation: 
\begin{align}
\left[\Sigma_{i}(\omega)\right]_{\alpha\beta}&=\langle\phi_{\bk i\alpha}|\omega+i0^+ - \left(\hat{\nabla}^2 - J(\hat{x}) -\mu\right) |\phi_{\bk i \beta}\rangle\nonumber \\
&-\left[G_{i}^{-1}(\mathbf{k},\omega)\right]_{\alpha\beta}.
\end{align} 
Note that the self-energy is momentum-independent in gRISB, \emph{i.e.}, $\Sigma_i(\bk,\omega)=\Sigma_i(\omega)$. 
The quasiparticle renormalization weight is determined from:
\begin{equation}
\mathbf{Z}=\left[1-\frac{\partial\text{Re}\boldsymbol{\Sigma}(\omega)}{\partial\omega}\Big|_{\omega\rightarrow0}\right]^{-1}.
\end{equation}
The expectation value of a generic local operator $\hat{O}$ is computed from the embedding wavefunction:
\begin{equation}
\langle \hat{O} \rangle = \langle \Phi_i | \hat{O}[{c_{i\alpha}^\dagger,c_{i\alpha}}] | \Phi_i \rangle
\,.
\end{equation}

\subsection{Implementation}

Our implementation closely follows the previous works~\cite{Lanata_2015_PRX,Lanata_2017_PRL}. We utilize Wien2k for the DFT part of the calculation~\cite{Wien2k}. The projector to the correlated orbitals is constructed from the atomic orbital modified from the density functional theory plus embedded dynamical mean-field theory (DFT+eDMFT) code~\cite{Haule_DFT+DMFT_2010,Yao2020_cygutz}. The temperature broadening method is utilized for the Brillouin zone integration with a broadening factor of 0.02 eV. The local density approximation functional (LDA) is utilized in our calculation. 
We use 5000 k-points and 2000 k-points for the NiO and SrVO$_{3}$, respectively, and the RK$_{\text{max}}$ is set to 7.
The energy window for constructing the low-energy Hubbard model is {[}-10 eV --- 10e V{]}.
The fully localized limit (FFL) is used as our double-counting scheme~\cite{FLL,Anisimov_1997_slater_condon,Haule_2015_exactDC}, where the nominal valence occupancy is set to 8 and 2 for NiO and SrVO$_{3}$, respectively.
We use the DMRG approach implemented in the block2 code to solve the ground-state wavefunction of $H^{\text{emb}}$~\cite{block2}.
For the DFT+DMFT calculations, we utilize the DFT+eDMFT code with the same parameter setting as in the DFT+gRISB calculations, which provides a consistent benchmark between the DFT+DMFT and DFT+gRISB methods.
The continuous-time Quantum Monte Carlo solver is utilized with $10^{9}$ Monte Carlo steps distributed over 200 CPUs, and the temperature is set to $100$ K. For both methods, we treat all five d-orbitals as correlated shells, and the interaction is of full Slater-Condon type~\cite{Anisimov_1993_slater_condon,Anisimov_1997_slater_condon}.

\begin{figure}[b]
\begin{centering}
\includegraphics[scale=0.4]{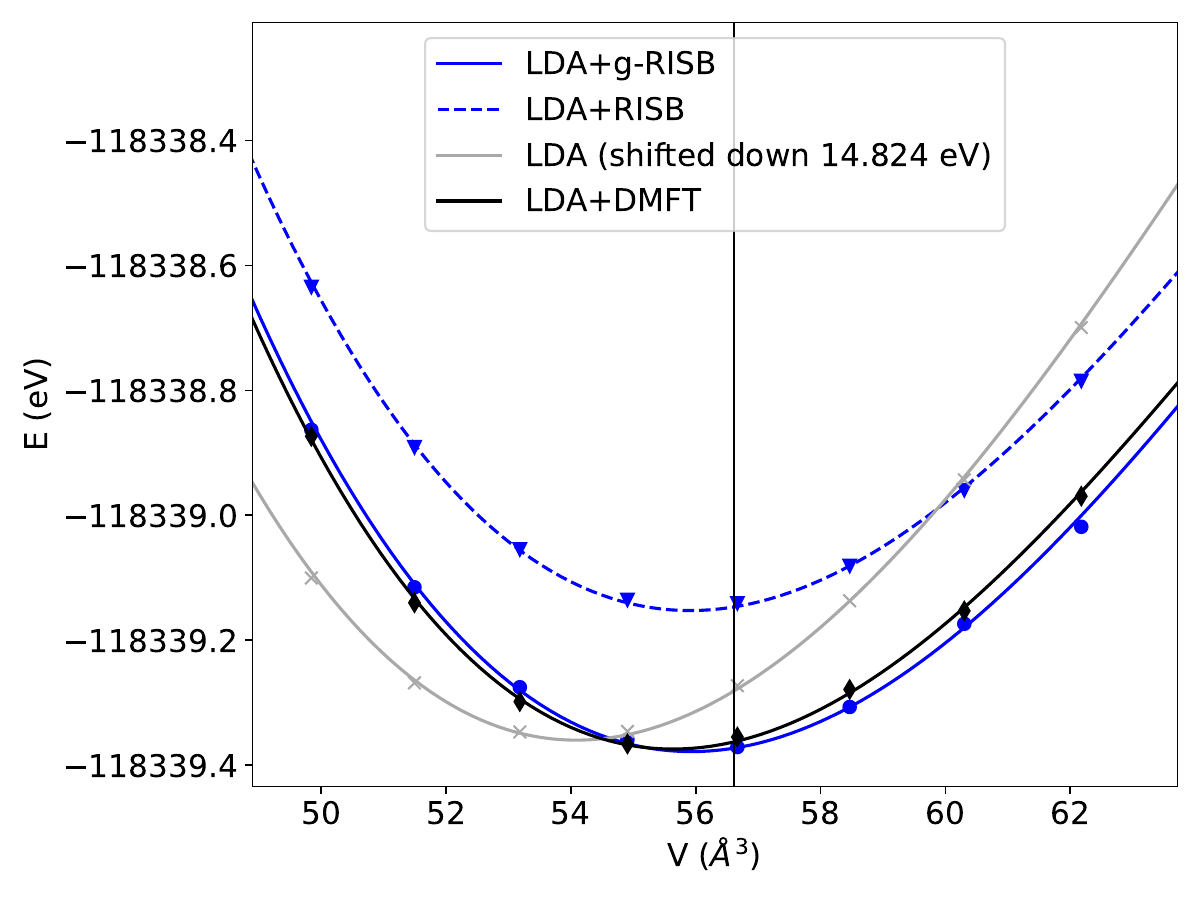}
\par\end{centering}
\caption{Calculated energy volume curve with LDA, LDA+RISB, LDA+gRISB, and LDA+DMFT for SrVO$_{3}$ with $U=10$ eV and $J=1$ eV.
The experimental equilibrium volume is 56.61 $\text{\AA}^{3}$~\cite{SVO_Vexp}. The temperature in DMFT is $T=100$ K. \label{fig:SVO_EV}}
\end{figure}

\begin{figure*}
\begin{raggedright}
\subfloat[LDA]{\begin{centering}
\includegraphics[scale=0.35]{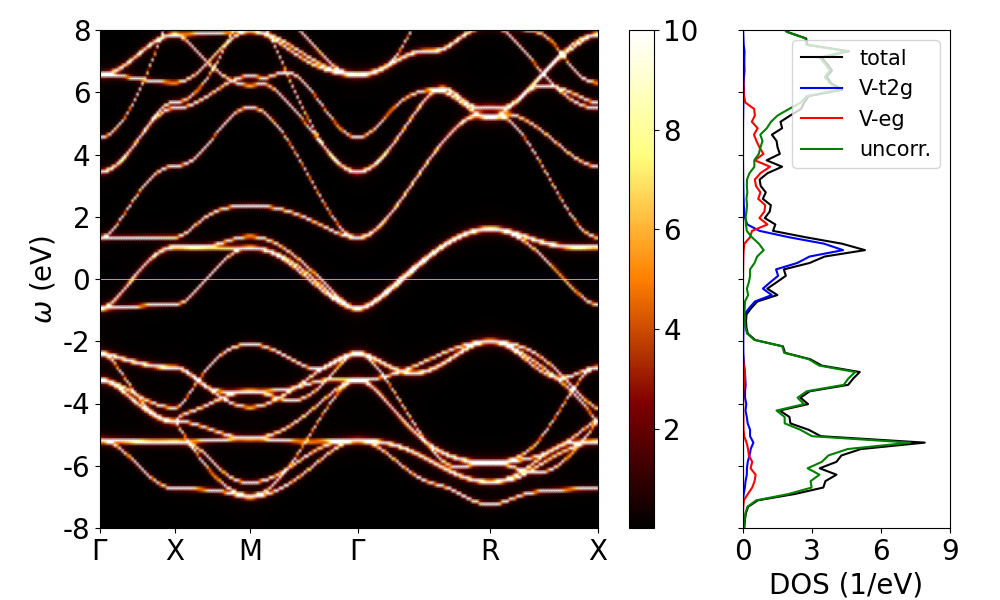}
\par\end{centering}
}\subfloat[LDA+RISB]{\begin{centering}
\includegraphics[scale=0.35]{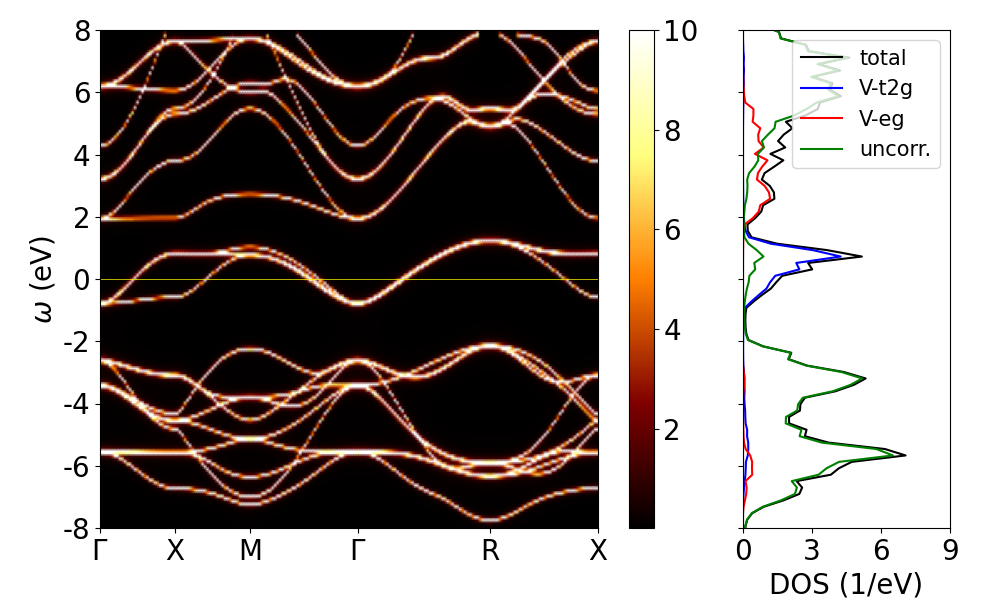}
\par\end{centering}
}
\par\end{raggedright}
\begin{centering}
\subfloat[LDA+gRISB]{\begin{centering}
\includegraphics[scale=0.35]{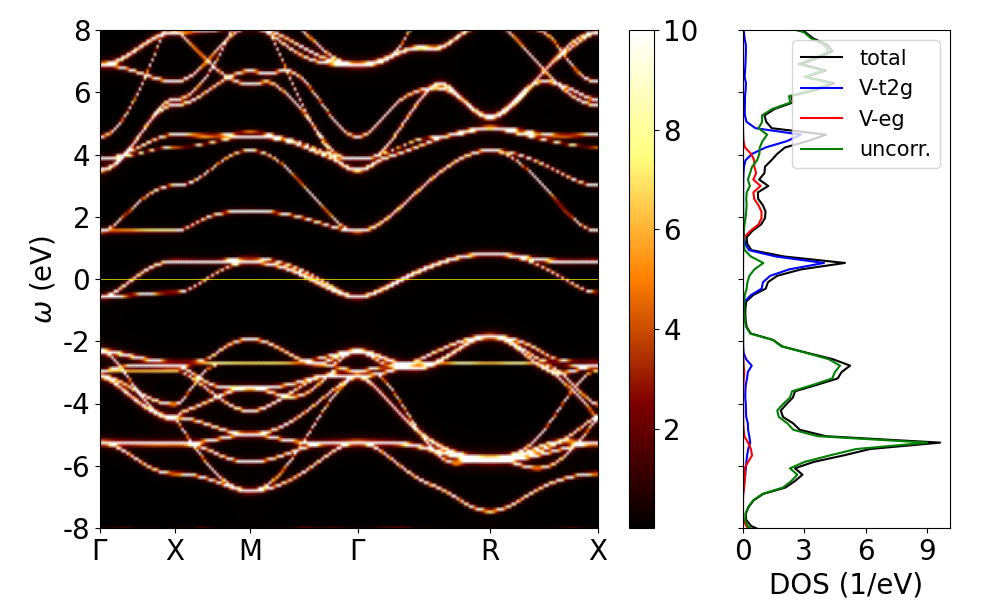}
\par\end{centering}
}\subfloat[LDA+DMFT]{\begin{centering}
\includegraphics[scale=0.35]{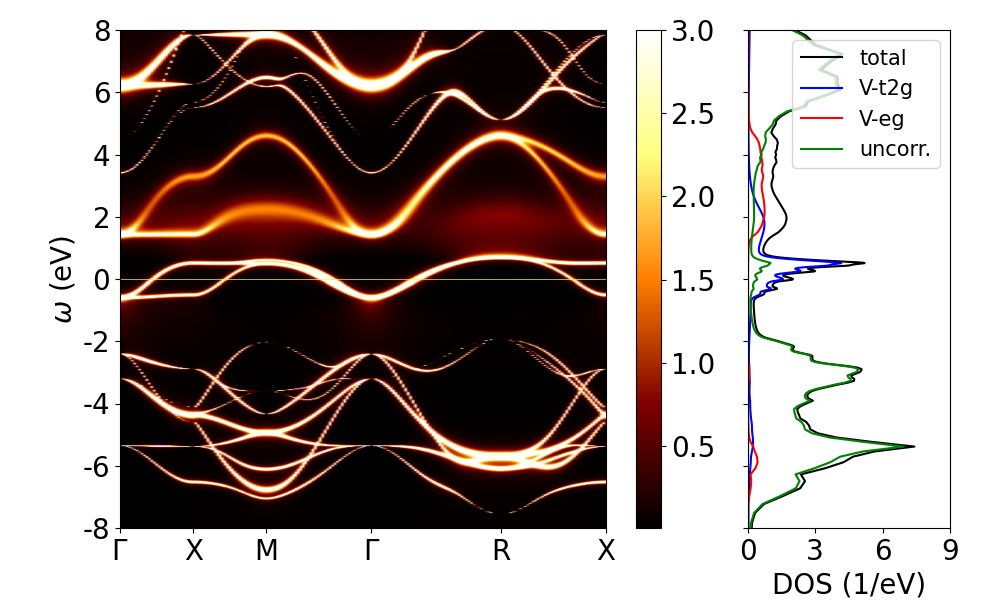}
\par\end{centering}
}
\par\end{centering}
\caption{The momentum-resolved spectral function $A(\mathbf{k},\omega)$ and the orbital-resolved density of state for SrVO$_3$ with the (a) LDA, (b) LDA+RISB, (c) LDA+gRISB, and (d) LDA+DMFT approaches at the experimental equilibrium volume. The Coulomb parameters are $U=10$ eV and $J=1$ eV, and the temperature in DMFT is $T=100$ K. \label{fig:Akw_DOS_SVO}}
\end{figure*}

The DFT+gRISB self-consistent equations are implemented as follows: (1) converge the DFT calculations to obtain the Kohn-Sham eigenvalue and eigenvectors, (2) construct the projector from the Kohn-Sham eigenvector and the local atomic orbitals, (3) solve the gRISB saddle-point equations Eqs. \ref{eq:sp3}-\ref{eq:sp9} with the Kohn-Sham eigenvalues and the projector, (4) use the gRISB saddle-point solution to compute the new charge density from Eq. \ref{eq:sp2}, (4) feedback the new charge density to DFT to update the new exchange-correlated potential (Eq. \ref{eq:sp1}) and go to step (1) until the charge density and total energy is converged.
In our calculations, we set the total energy convergence criteria to $10^{-5}$ eV and the charge convergence criteria to $10^{-3}$.

\section{Results}

\subsection{Applications to SrVO$_{3}$}

In this section, we apply DFT+gRISB to SrVO$_3$ and investigate its total energy and electronic structures. This material has been studied extensively in the past decades and serves as an ideal material for benchmarking new approaches~\cite{fujimori_oxides_1992,svo_sco_weight,SVO_arpes,Liebsch_2003,Pavarini_2004,Sekiyama_2004,Yoshida2005,Nekrasov2006,Tomczak_2012,Taranto_2013,Tomczak_2014,Haule_2014_nnom,Haule_free_energy_SVO_2015,Haule_2015_exactDC,Zhong_2015_svo,Birol_2019,Aichhorn_2021,Tomczak_2021}. It has a cubic perovskite structure, and the main active orbitals around the Fermi level are in the V-$d_{t_{2g}}$ shell. The correlation effect is essential in SrVO$_3$, leading to significant renormalization of the bandwidth near the Fermi level.

We first discuss the total energy of the DFT+gRISB. Figure \ref{fig:SVO_EV} summarized the total energy of LDA, LDA+RISB, LDA+gRISB, and LDA+DMFT as a function of the unit cell volume.
First, we reproduce the known fact that LDA underestimates the equilibrium volume at $54$ $\text{\AA}$, while the experiment observed value is $56.61\ \text{\AA}$. 
The LDA+RISB improves the equilibrium volume to $56\ \text{\AA}$ towards the experimental value, but the total energy is not consistent with LDA+DMFT at the quantitative level. On the other hand, LDA+gRISB with 15 bath orbitals significantly improves the total energy, in quantitative agreement with the LDA+DMFT results.
The equilibrium volume of LDA+gRISB and LDA+DMFT is
$56\ \text{\AA}$ and $55.8\ \text{\AA}$, respectively. 

We now discuss the electronic structure of SrVO$_{3}$. Figure~\ref{fig:Akw_DOS_SVO} are the momentum-resolved spectral function along the high-symmetry points and the orbital resolved density of state calculated from the LDA, LDA+RISB, LDA+gRISB, and LDA+DMFT approaches at the experimental equilibrium volume.
The main characters around the Fermi level are the Vanadium's $t_{2g}$ orbitals. For the LDA calculation, the bandwidth of the $t_{2g}$ bands is around 2 eV. 
When including the electronic correlation effects at the LDA+RISB level, we observe slight renormalization of the bandwidth by a factor around $0.8$, which is inconsistent with the LDA+DMFT, where the renormalization factor is around $0.5$.
Moreover, the LDA+RISB electronic structure is almost identical to LDA, implying a weakly correlated metal that is inconsistent with LDA+DMFT. This inconsistency can be remedied by utilizing LDA+gRISB with 15 bath orbitals shown in Fig. \ref{fig:Akw_DOS_SVO}(c). 
In the LDA+gRISB results, the $t_{2g}$ bands are renormalized by a factor of $0.5$ in agreement with LDA+DMFT.
Moreover, the electronic structure closely resembles LDA+DMFT, except for the upper Hubbard band of the $t_{2g}$ orbitals is approximated by coherent bands in LDA+gRISB, which is an established feature of gRISB~\cite{gRISB_2017,Lee_one-orbital_2023,Lee_multiorb_2023}.

\begin{figure}[t]
\begin{centering}
\includegraphics[scale=0.4]{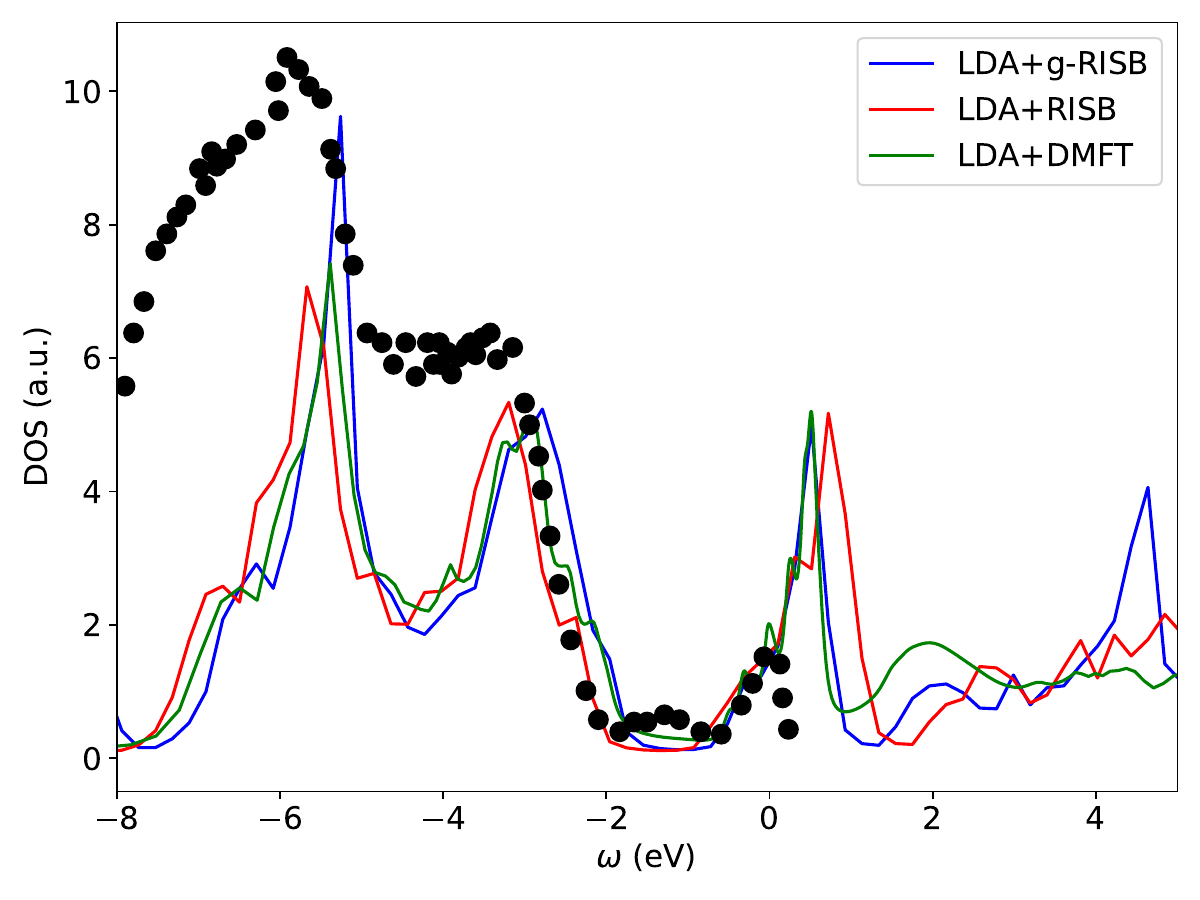}
\par\end{centering}
\caption{Comparison of the DFT+RISB, DFT+gRISB, and DFT+DMFT SrVO$_{3}$ total density of states with the Photoemission spectroscopy (the filled circles)~\cite{SVO_arpes}. The Coulomb interaction parameters are $U=10$ eV and $J=1$ eV. The temperature in DMFT is $T=100$ K. \label{fig:DOS_XPS_SVO}}
\end{figure}

The total density of states calculated from LDA+RISB, LDA+gRISB, and LDA+DMFT at the equilibrium volume is shown in Fig. \ref{fig:DOS_XPS_SVO}, which are compared with the photoemission experiment~\cite{Allen1984}.
The LDA+gRISB and LDA+DMFT density of states around the Fermi level are in good agreement with each other, while the upper Hubbard band in LDA+gRISB shows a coherent feature. 
All three methods capture the gap between -2 eV and 0.5 eV and the low energy peaks, mainly attributed to the uncorrelated O and Sr atoms.
Finally, we show the orbital-resolved quasiparticle weight $Z$ and occupancy in Table~\ref{tab:Z_n_SVO}.
The LDA+gRISB and LDA+DMFT values are in quantitative agreement. On the other hand, the LDA+RISB captures reliable occupancy, but the quasiparticle weight is overestimated.

\begin{table}[ht]
\begin{centering}
\begin{tabular}{|c|c|c|c|c|}
\hline 
 & LDA & LDA+RISB & LDA+gRISB & LDA+DMFT\tabularnewline
\hline 
\hline 
$Z_{t2g}$ & 1 & 0.78 & 0.53 & 0.51\tabularnewline
\hline 
$Z_{eg}$ & 1 & 0.89 & 0.75 & 0.78\tabularnewline
\hline 
$n_{t2g}$ & 1.68 & 1.56 & 1.53 & 1.51\tabularnewline
\hline 
$n_{eg}$ & 0.80 & 0.64 & 0.66 & 0.66\tabularnewline
\hline 
\end{tabular}
\par\end{centering}
\caption{Quasiparticle weight $Z$ and occupancy $n$ for SrVO$_3$ calculated from LDA, LDA+RISB, and LDA+DMFT with Coulomb interaction parameters $U=10$ eV and $J=1$ eV at the experimental equilibrium volume.
\label{tab:Z_n_SVO}}
\end{table}

\subsection{Applications to NiO}

We now apply DFT+gRISB to NiO, which is a prototypical charge-transfer insulator where the Ni-$d_{e_{g}}$ orbitals hybridize with O-$p$ to form the so-called Zhang-Rice band between the lower and the upper Hubbard band\citep{Zaanen_Sawatsky_1985,ZR_1988,Kunes_2007}. The DFT+DMFT approach has been applied to NiO and reveals further insight into electronic structure of this charge-transfer insulator as well as its properties with external pressure, doping, and the surface effects~\citep{Kunes_2007,Kunes_2007_PRB,NiO_Zhange_2019,Leonov_2016,Mandal_2019,Mandal_2019_prb,Leonov_2020,Karolak_2010,Zhu_2020}.
In this work, we focus on the paramagnetic phase of NiO and compare its total energy and electronic structure with the DFT+DMFT results.

\begin{figure}[t]
\begin{centering}
\includegraphics[scale=0.4]{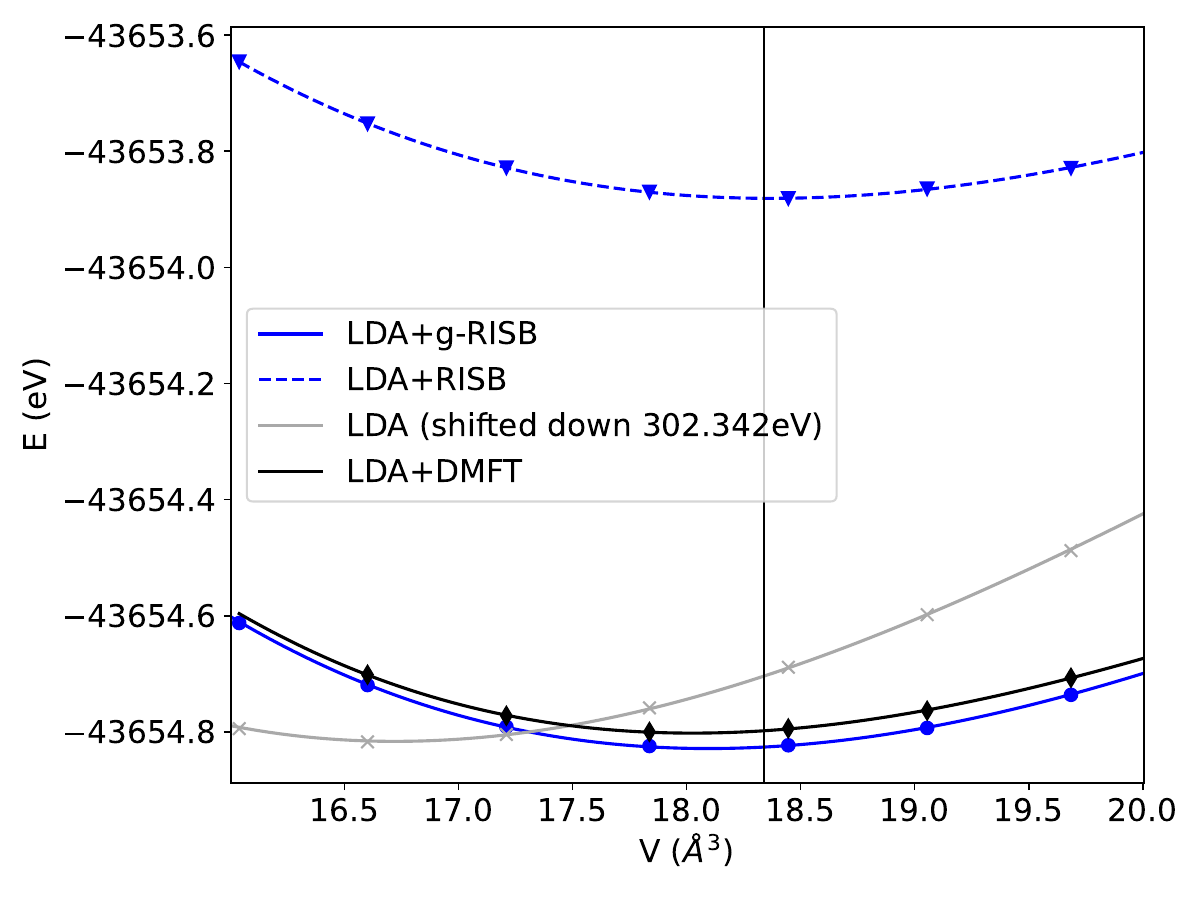}
\par\end{centering}
\caption{Calculated energy volume curve with LDA, LDA+RISB, LDA+gRISB, and LDA+DMFT for NiO with Coulomb interaction parameters $U=10$ eV and $J=1$ eV. The experimental equilibrium volume is 18.34 $\text{\ensuremath{\mathring{A}^{3}}}$~\cite{NiO_Vexp}.  The temperature in DMFT is $T=100$ K.\label{fig:NiO_EV}}
\end{figure}

\begin{figure*}
\begin{centering}
\subfloat[LDA]{\begin{centering}
\includegraphics[scale=0.35]{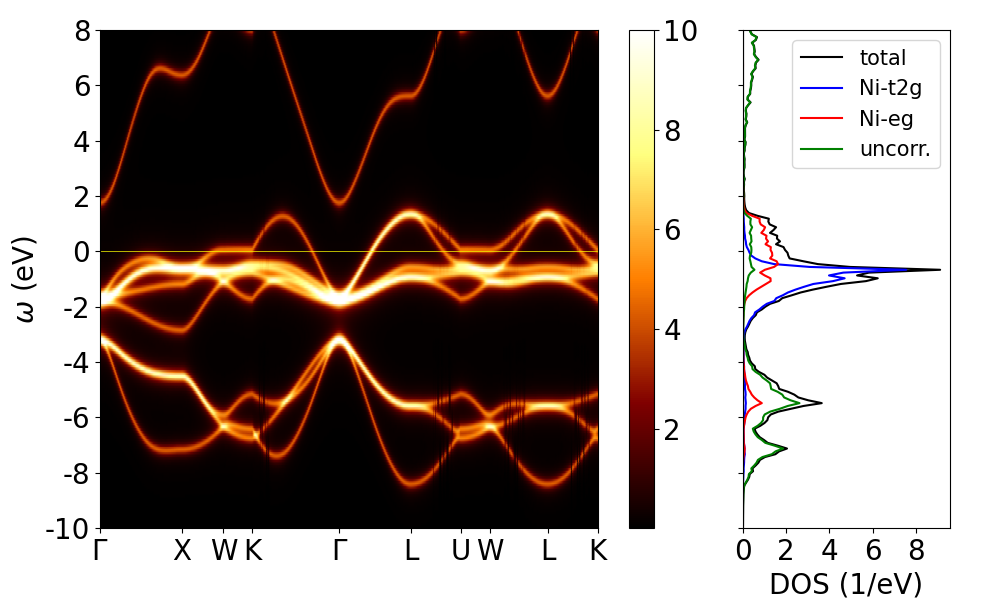} 
\par\end{centering}
}\subfloat[LDA+RISB]{\begin{centering}
\includegraphics[scale=0.35]{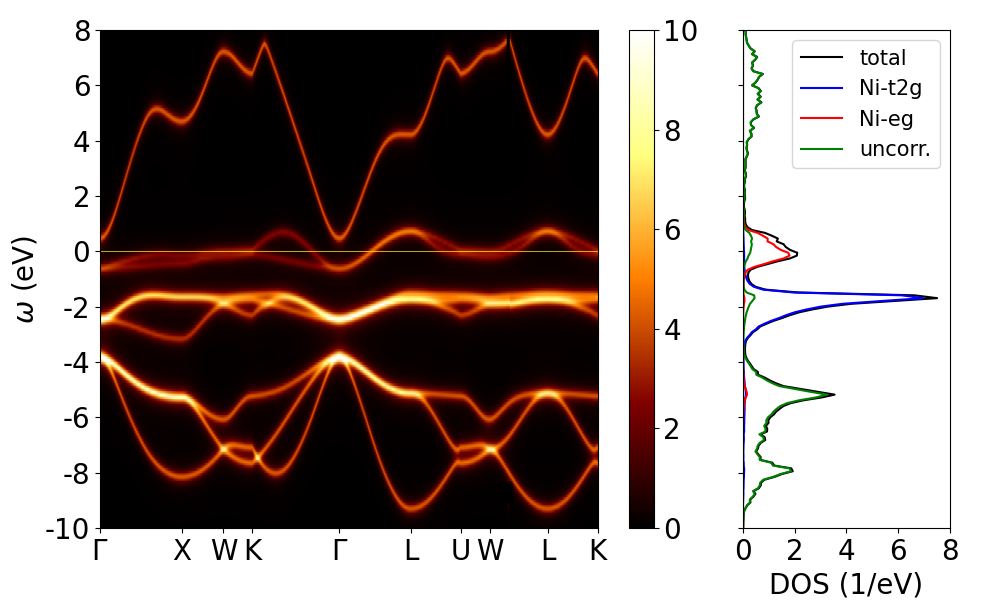} 
\par\end{centering}
}
\par\end{centering}
\begin{centering}
\subfloat[LDA+gRISB]{\begin{centering}
\includegraphics[scale=0.35]{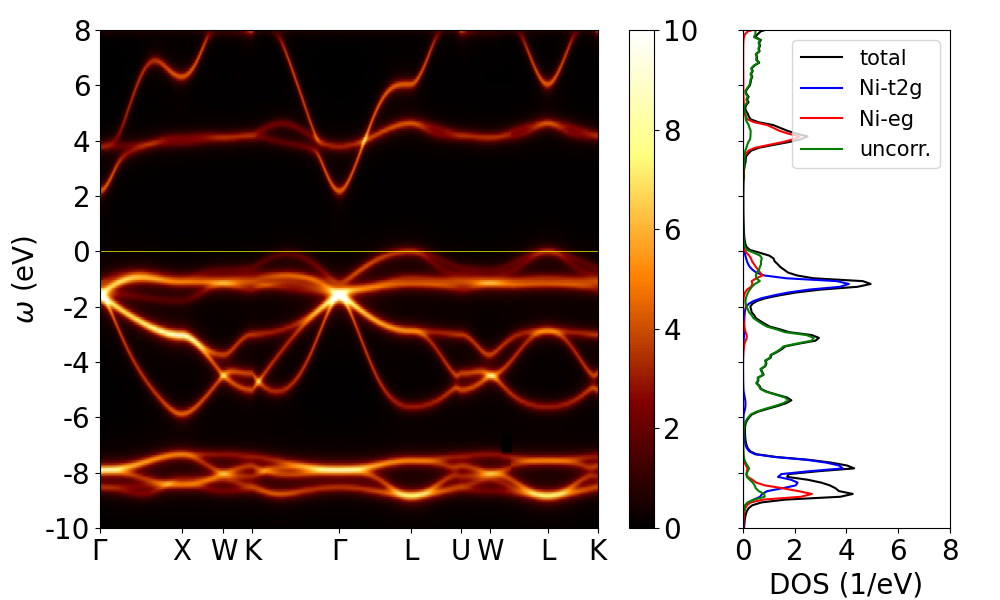}
\par\end{centering}
}\subfloat[LDA+DMFT]{\begin{centering}
\includegraphics[scale=0.35]{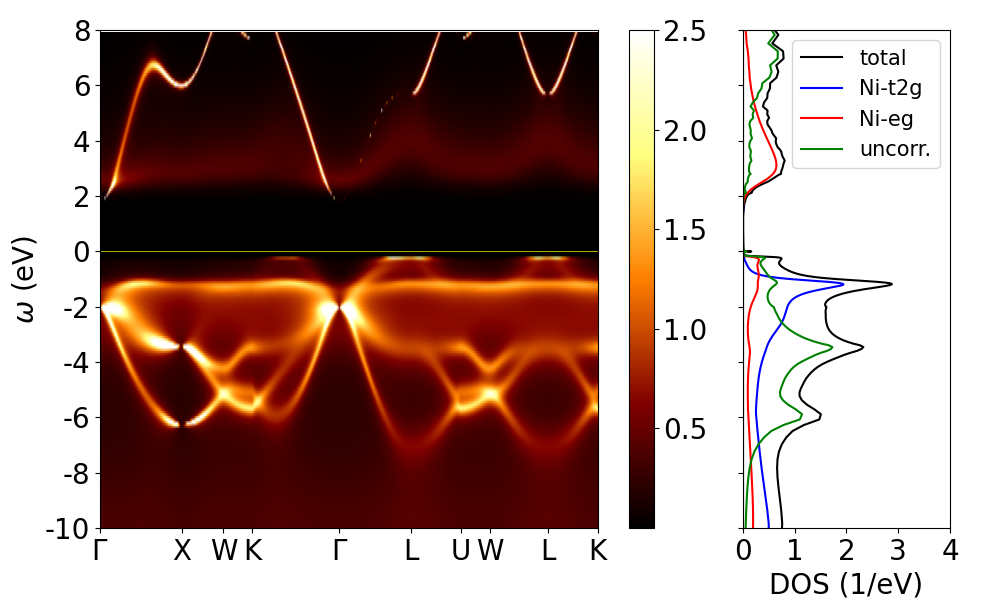}
\par\end{centering}
}
\par\end{centering}
\caption{The momentum-resolved spectral function $A(\mathbf{k},\omega)$ and the orbital-resolved density of state for NiO with the (a) LDA, (b) LDA+RISB, (c) LDA+gRISB, and (d) LDA+DMFT approaches at the experimental equilibrium volume. The Coulomb parameters are $U=10$ eV and $J=1$ eV, and the temperature in DMFT is $T=100$ K. \label{fig:Akw_DOS_NiO}}
\end{figure*}

The total energy as a function of the unit cell volume is shown in Fig. \ref{fig:NiO_EV} for LDA, LDA+RISB, LDA+DMFT, and LDA+DMFT. The LDA significantly underestimates the unit cell volume around 16.6 $\text{\AA}$ and fails to capture the experimentally observed insulating behavior, which is a well-known feature.
The LDA+RISB also fails to produce a Mott insulating solution with realistic Coulomb parameters, $U=10$ eV and $J=1$ eV~\citep{NiO_UJ_1991,Kunes_2007}, utilized in this work. Therefore, its total energy has a large 1 eV discrepancy compared to LDA+DMFT.
On the other hand, LDA+gRISB captures the charge transfer insulator behavior, significantly improving the total energy to a quantitative agreement with the LDA+DMFT values, with a difference of around 10 meV. This difference can be attributed to the finite temperature effect $T=100$ K utilized in the LDA+DMFT calculations.

The DFT bandstructure and density of states are shown in Fig. \ref{fig:Akw_DOS_NiO}(a).
Without breaking the spin symmetry, DFT predicts a metallic solution, which is a known problem in DFT for transition metal oxides \citep{Mattheiss1972}.
The bands around the Fermi level contain the O-$p$ and Ni-$d_{e_{g}}$ orbital. The Ni-$t_{2g}$orbitals are located below the Fermi level and are almost completely filled.

Next, we show the DFT+RISB momentum-resolved spectral function and density of states in Fig. \ref{fig:Akw_DOS_NiO}(b).
We utilize the constrained LDA Coulomb parameters $U=10$ eV and $J=1$ eV in our simulations \citep{NiO_UJ_1991,Kunes_2007}. In DFT+RISB, the correlation effects only slightly renormalized the bands around the Fermi level with $Z_{d_{eg}}=0.6$ and $Z_{d_{t2g}}=0.7$ and the system is far from the metal-insulator transition.

The DFT+gRISB momentum-resolved spectral function and density of states are shown in Fig. \ref{fig:Akw_DOS_NiO}(c). Here, we use 17 bath orbitals in our DFT+gRISB calculations. The density of states shown in Fig. \ref{fig:Akw_DOS_NiO}(c) demonstrate that DFT+gRISB accurately captures the charge-transfer insulating behavior in the density of state, where the Hubbard bands are opened in the Ni-$d_{e_{g}}$ orbitals and the Zhang-Rice band is observed around $-1$ eV with strong $O$-$p$ and $Ni$-$d$ hybridization, in good agreement with the DFT+DMFT results shown in Fig. \ref{fig:Akw_DOS_NiO}(d) and the previous studies \citep{Kunes_2007,Kunes_2007_PRB,NiO_Zhange_2019,Leonov_2016,Leonov_2020,Karolak_2010,Zhu_2020}.
Moreover, the DFT+gRISB momentum-resolved spectral functions $A(\mathbf{k},\omega)$ captures reliably the dispersive excitations compared to DFT+DMFT, except the incoherent broadening features, which cannot be described within the gRISB framework.

\begin{figure}[t]
\begin{centering}
\includegraphics[scale=0.4]{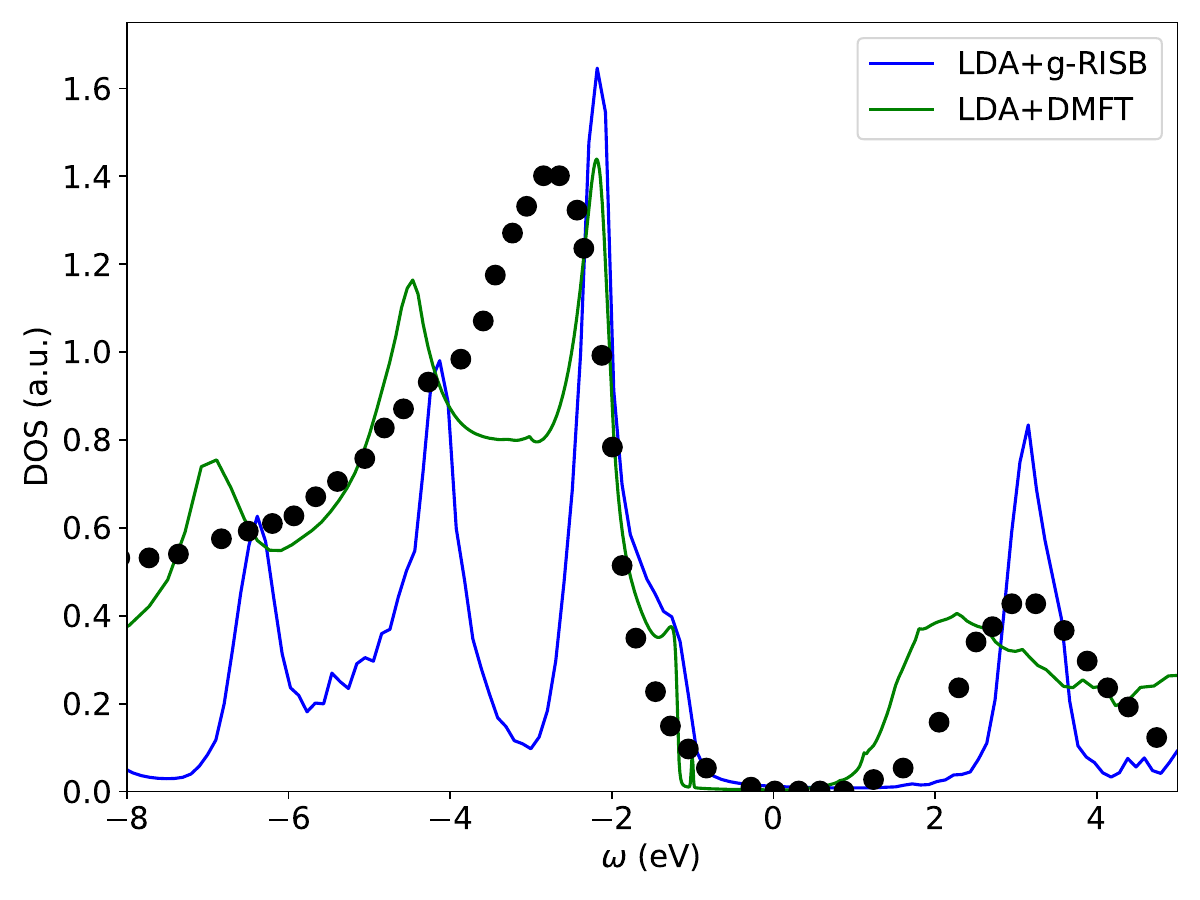}
\par\end{centering}
\caption{Comparison of the DFT+gRISB and DFT+DMFT NiO total density of states with the Photoemission and bremsstrahlung isochromat spectroscopy (the filled circles)~\cite{Allen1984}. The Coulomb interaction parameters are $U=10$ eV and $J=1$ eV. The temperature in DMFT is $T=100$ K.
\label{fig:DOS_XPS_NiO}}
\end{figure}

In Fig.~\ref{fig:DOS_XPS_NiO}, we compare the LDA+gRISB total density of states with LDA+DMFT and the photoemission bremsstrahlung-isochromat-spectroscopy (XPS/BIS)~\citep{Allen1984}.
Our LDA+gRISB density of states captures the main features in the XPS/BIS spectrum, where the band gap is about $4$ eV, and the heights of the peaks on the band edges are reasonably captured. On the other hand, LDA+DMFT has band gap around 2 eV, smaller than the experimental band gap.

Finally, the orbital-resolved occupancy of different approaches is shown in Table~\ref{tab:n_NiO}. The LDA+gRISB's occupancy is in good agreement with LDA+DMFT and improves the LDA values. On the other hand, although LDA+RISB fails to capture the charge-transfer insulating solution, its occupancy is identical to the LDA+gRISB values and close to the LDA+DMFT values.

\begin{table}[t]
\begin{centering}
\begin{tabular}{|c|c|c|c|c|}
\hline 
 & LDA & LDA+RISB & LDA+gRISB & LDA+DMFT\tabularnewline
\hline 
\hline 
$n_{eg}$ & 2.59 & 2.14 & 2.15 & 2.15\tabularnewline
\hline 
$n_{t2g}$ & 5.89 & 5.99 & 5.99 & 5.94\tabularnewline
\hline 
\end{tabular}
\par\end{centering}
\caption{The occupancy $n$ for NiO calculated from LDA, LDA+RISB, and LDA+DMFT with Coulomb interaction parameters $U=10$ eV and $J=1$ eV at the experimental equilibrium volume.
\label{tab:n_NiO}}
\end{table}

\section{Conclusions}

We present a charge-self-consistent DFT+gRISB approach to correlated materials and demonstrate its performance on two prototypical materials, SrVO$_{3}$ and NiO, representing the correlated metals and charge transfer insulators.
For SrVO$_{3}$, we show that DFT+gRISB reliably captures the total energy and effective mass compared to the experiment and DFT+DMFT values, significantly improving the original DFT+RISB approach.
Furthermore, we show that DFT+gRISB provides a more accurate description of the electronic band structure for strongly correlated materials with a narrow quasiparticle peak and Hubbard bands compared to DFT+RISB.
For NiO, DFT+gRISB captures the charge transfer insulating behavior, with the Zhang-Rice band forming between the lower and upper Hubbard bands, significantly improving the DFT+RISB results, which falsely predict a metallic state. The total density of states is in reasonable agreement with the photoemission spectrum.
Moreover, our work demonstrates the applicability of DMRG as an impurity solver within the DFT+gRISB framework to reliably simulate correlated full five d-orbital systems.

Future work will extend the DFT+gRISB framework to study the two-particle response functions and interaction vertices~\citep{von_Oelsen_2011_PRL,Lee_PRX_2021}, non-equilibrium dynamics~\citep{Schiro2011,Behrmann_2013,Behrmann_2016,Guerci_2023}, and to incorporate gRISB with Wannier-orbital-based projectors and other DFT frameworks and interfaces~\citep{Lecherman_2006,Aichhorn_2009,Aichhorn_2011,triqs,dfttools,Sigh_2021,Beck_2022,Grechnev_2007,DiMarco_2009,granas_2012,Dcore,Adler2024,ComDMFT1,ComDMFT2}.

\begin{acknowledgments}
T.-H.L acknowledges discussions with Huanchen Zhai on the block2 DMRG solver. T.-H.L,  X.S., and G.K. were supported by the U.S. Department of Energy, Office of Science, Office of Advanced Scientific Computing Research and Office of Basic Energy Sciences, Scientific Discovery through Advanced Computing (SciDAC) program under Award Number DE-SC0022198.
C.M. and  R.A., was supported by the U.S. Department of Energy, Office of Basic Energy Sciences as part of the Computation Material Science Program.
T.-H.L. gratefully acknowledges funding from the Ministry of Science and Technology of Taiwan under Grant No. NSTC 112-2112-
M-194-007-MY3.
N.L. gratefully acknowledges funding from the Simons Foundation (Grant No. 1030691, N.L.).
The part of the work by Y.Y. was supported by the U.S. Department of Energy (DOE), Office of Science, Basic Energy Sciences, Materials Science and Engineering Division, including the grant of computer time at the National Energy Research Scientific Computing Center (NERSC) in Berkeley, California. This part of research was performed at the Ames National Laboratory, which is operated for the U.S. DOE by Iowa State University under Contract No. DE-AC02-07CH11358.
\end{acknowledgments}

\bibliographystyle{apsrev}
\bibliography{ref}

\end{document}